# Jumps in entropy and magnetic susceptibility at the valence and spin-state transition in a cobalt oxide

V. Hardy, F. Guillou, and Y. Bréard

Laboratoire CRISMAT, ENSICAEN, UMR 6508 CNRS, 6 Boulevard du Maréchal Juin, 14050 Caen Cedex, France.

**Abstract**

A wide family of cobalt oxides of formulation $(Pr_{1-y}Ln_y)_{1-x}Ca_xCoO_3$ (Ln being a lanthanide) exhibits a coupled valence and spin-state transition (VSST) at a temperature $T^*$, which involves two concomitant modifications: (i) a change in the spin state of $Co^{3+}$ from low-spin ($T < T^*$) to a higher spin-state ($T > T^*$), and (ii) a change in the valence state of Pr, from a mixed $Pr^{4+}/Pr^{3+}$ state ($T < T^*$) to a purely trivalent state ($T > T^*$), accompanied by an equivalent charge transfer within the $Co^{3+}/Co^{4+}$ subsystem. In the present paper, the VSST taking place in $(Pr_{0.7}Sm_{0.3})_{0.7}Ca_{0.3}CoO_3$ at $T^* \sim 90$ K is investigated by magnetization and heat capacity measurements. First, we quantitatively characterized the jumps in magnetic susceptibility ($\chi$) and entropy ($S$) around $T^*$. Then, these values were compared to those calculated as a function of the variations in the population of the different cationic species involved in the VSST. X-ray absorption spectroscopy experiments recently showed that the higher spin state above $T^*$ should be regarded as an inhomogeneous mixture between low-spin (LS) and high-spin (HS) states. In the frame of this description, we demonstrate that the jumps in both $\chi$ and $S$ can be associated to the *same change* in the $Co^{3+}$ HS content around $T^*$. This result lends further support to the relevance of the LS/HS picture for the VSST, challenging the currently dominant interpretation based on the occurrence of an intermediate-spin (IS) state of $Co^{3+}$ above $T^*$.

# 1. Introduction

In 2002, Tsubouchi *et al.* reported a new type of spin-state transition (SST) in cobalt oxides:[1] in well oxygenated $Pr_{0.5}Ca_{0.5}CoO_3$ materials, they observed the concomitant occurrence of sharp jumps in magnetization, entropy and unit-cell volume, at $T^* \sim 90$ K. In the following years, the same type of first-order like transition was found in other perovskite cobaltites of formulation $(Pr_{1-y}Ln_y)_{1-x}Ca_xCoO_3$, where Ln =Sm, Eu, Gd, Tb or Y, while $x$ and $y$ are approximately within the ranges 0.2-0.5 and 0-0.3, respectively.[1-8] The interpretation of this transition was initially ascribed to a pure SST within the octahedrally coordinated $Co^{3+}$, shifting from a low-spin (LS, $t_{2g}^6 e_g^0$) state below $T^*$ to an intermediate-spin (IS, $t_{2g}^5 e_g^1$) state above $T^*$.[1-3] Starting from 2010 however, it was shown that the stabilization of a mixed $Pr^{3+}/Pr^{4+}$ valence at low temperatures was also involved in this transition.[8,9-13] In addition to the SST transition affecting the $Co^{3+}$, a $Pr^{4+}$-to-$Pr^{3+}$ transformation takes place when crossing $T^*$ upon warming, being counterbalanced by a corresponding change from $Co^{3+}$ to $Co^{4+}$.[8,10-12] The transition at $T^*$ in these compounds can thus be regarded as a coupled valence and spin-state transition (VSST).

At the present time, the picture which is the most commonly accepted about the VSST in $(Pr_{1-y}Ln_y)_{1-x}Ca_xCoO_3$ is to consider that the $Co^{4+}$ are LS, while the $Co^{3+}$ undergo a transition from LS (below $T^*$) to IS (above $T^*$).[1,3-5,7,8,13] In a recent x-ray absorption spectroscopy (XAS) study on $Pr_{0.5}Ca_{0.5}CoO_3$, however, Herrero *et al.* showed that the spin state of $Co^{3+}$ above $T^*$ can be equally well described by an inhomogeneous mixture between low-spin and high-spin (HS, $t_{2g}^4 e_g^2$) states.[14] Still more recently, another XAS study –performed in $(Pr_{0.7}Sm_{0.3})_{0.7}Ca_{0.3}CoO_3$– led us to consider that the achievement of a mixed LS/HS state above $T^*$ is the most likely option.[15]

It is known that magnetization and heat capacity data can provide valuable complementary information, when dealing with the spin-state issue in cobalt oxides.[16-20] The goal of the present study is to relate the jump in $\chi(T)$ and the peak in $C(T)$, that are observed at $T^*$ in $(Pr_{0.7}Sm_{0.3})_{0.7}Ca_{0.3}CoO_3$, to the various changes (valence and spin-state) expected to take place at the VSST. We include in this analysis the results of our XAS study, which quantified the amount of $Pr^{4+}$ stabilized below $T^*$ in $(Pr_{0.7}Sm_{0.3})_{0.7}Ca_{0.3}CoO_3$ ( i.e., 0.13 per f.u. at 10 K, corresponding to a ratio $Pr^{4+}/Pr^{3+} \sim 0.36$).[15] In terms of Co spin-states, this XAS study showed that $Co^{3+}$ are initially purely LS at $T<<T*$ and undergo a partial excitation to HS upon warming across $T*$, leading to a mixed LS/HS state; it also indicated that the $Co^{4+}$ are in an IS state,[21] instead of LS as usually considered in the literature.[15] It must be emphasized that the VSST was systematically found to develop in the paramagnetic regime; indications of strong ferromagnetic correlations and possible short-range ordering (likely related to $Co^{4+}$) were only reported at much lower temperatures (typically at $T \leq 5$ K).[1,3,8]

## 2. Experimental details

Polycrystalline samples of $(Pr_{0.7}Sm_{0.3})_{0.7}Ca_{0.3}CoO_3$ were prepared by solid-state reaction, using stoichiometric proportions of high purity $Pr_6O_{11}$, $Sm_2O_3$, CaO and $Co_3O_4$ precursors. The samples were first sintered at 1200 °C in flowing oxygen for 36 h, and then annealed in high-pressure (130 bar) $O_2$ atmosphere for 48 h at 600 °C. It was previously checked by various techniques (iodometric titrations, thermo-gravimetric analysis, and Rietveld refinement of neutron diffraction data)[1,2,12] that such a procedure leads to full oxygen stoichiometry (within the experimental uncertainties) in the related compounds $Pr_{1-x}Ca_xCoO_3$ with $x\sim0.5$. Powder x-ray diffraction showed the $(Pr_{0.7}Sm_{0.3})_{0.7}Ca_{0.3}CoO_3$ samples are single-phase, with orthorhombic symmetry (space group P$nma$) and parameters [$a = 5.3461(8)$ Å, $b = 7.5518(8)$ Å, and $c = 5.3499(7)$ Å] leading to a unit-cell volume ($\approx 215.99$ Å$^3$) in agreement with the literature.[4]

Magnetization measurements were carried out with a superconducting quantum interference device magnetometer (MPMS, Quantum Design). Isothermal magnetization curves at selected temperatures demonstrated that the magnetization ($M$) varies linearly with the magnetic field ($H$) for temperature higher than $\sim 20$ K. Accordingly, an isofield magnetization curve was measured in a field of 1 T and the dc susceptibility $\chi$ ($T > 20$ K) was derived from the ratio $M/H$. The sample was first cooled in 1 T down to 5 K, and the data was then recorded upon warming (field-cooled warming procedure). Zero-field resistivity was measured upon warming by using a standard four-probe method in a "physical properties measurements system" (PPMS, Quantum Design). Heat capacity measurements were carried out in the same device by using a relaxation method. We combined measurements derived from the built-in $2\tau$ model (outside the transition region) and the result of a single-pulse method around $T*$ (85 K $\leq T \leq$ 93 K).[22] The latter method is required owing to the sharpness of the peak and the hysteretic effects resulting from the first-order nature of the transition.[3] To avoid alteration of the peak shape, we analyzed the relaxation branch (no heating power) of the heating pulse spanning the first-order part of the transition.

## 3. Results and discussion

Figure 1 shows that the VSST manifests itself by (a) a sharp peak on *C(T)*; (ii) an upward shift in $\chi(T)$ as *T* is increased; (iii) and a sharp decrease in resistivity. The inflection points in (b) and (c) are close to ~ 92.5 K while the peak in (a) is centered at ~ 89.5 K. Even though uncertainties in the thermometry of the various experimental setups can play a role, the observed scatter in *T\** is mainly ascribable to a hysteretic effect. Indeed, the lower value obtained from *C(T)* actually corresponds to a portion of the curve that was recorded upon cooling (relaxation branch in the single-pulse method), whereas both $\chi(T)$ and $\rho(T)$ correspond to continuous heating. In the literature on VSST, typical values of the thermal hysteresis were found to lie between 2 and 4 K.[1,6,9]

*3.1 Position of the problem*

In the present compound –which can also be written as $Pr_{0.49}Sm_{0.21}Ca_{0.3}CoO_3$– there are five cations whose content can change around the VSST: $Pr^{3+}$, $Pr^{4+}$, $Co^{4+}$ IS, $Co^{3+}$ LS and $Co^{3+}$ HS. These populations must obey the following relationships at all temperatures (amounts per f.u. are written in square brackets): (i) $[Pr^{4+}] + [Pr^{3+}] = 0.49$; (ii) $[Co^{4+} \text{ IS}] + [Co^{3+} \text{ LS}] + [Co^{3+} \text{ HS}] = 1$; (iii) $3[Pr^{3+}] + 4[Pr^{4+}] + (3\times 0.21) + (2\times 0.3) + 4[Co^{4+} \text{ IS}] + 3([Co^{3+} \text{ LS}] + [Co^{3+} \text{ HS}]) = 6$. There are thus only two independent variables, that we chose to be $[Pr^{4+}]$ and $[Co^{3+}\text{HS}]$. The other three parameters are given by

$$[Pr^{3+}] = 0.49 - [Pr^{4+}] ,  \quad (1.a)$$
$$[Co^{4+} \text{ IS}] = 0.3 - [Pr^{4+}] , \quad (1.b)$$
$$[Co^{3+} \text{ LS}] = 0.7 + [Pr^{4+}] - [Co^{3+} \text{ HS}]. \quad (1.c)$$

Another important issue is the "choice" of the temperature range used for the experimental determinations and calculations of the jumps in $\chi(T)$ and *S(T)*. The fact is there is no experimental signatures marking precisely and unambiguously the boundaries of the transition (see Fig. 1). Even for $\chi(T)$ –and despite the sharpness of the variation around *T\**– the rounding of the curve at both sides of the transition impedes any univocal definition of a "susceptibility jump". We thus preferred to address the variations in $\chi(T)$ and $S(T)$ between two –somewhat arbitrary– temperatures clearly located at each side of the transition. Considering linear regimes at each side of the transition in Fig. 1(a), one observes that the $C(T)$ peak roughly starts at 70 K and ends at 110 K. An upper boundary at 110 K is also well consistent with the $\chi(T)$ data, in that it corresponds to the beginning of a clear deviation from the Curie-Weiss regime present at $T > T^*$ [solid line in Fig. 1(b)]. Accordingly, we will consider the variations in $\chi$ and $S$ between a low-$T$ boundary at 70 K (noted LT) and a high-$T$ boundary at 110 K (noted HT), symmetrically distributed around $T^*$. We emphasize that both LT and HT must just be regarded as *reference* temperatures allowing to compare experimental and calculated values of $\Delta\chi$ and $\Delta S$.

The literature on VSST shows that the valence and spin-state change exhibits a sudden acceleration at $T^*$ (yielding a first-order character to this transition), but it is actually superimposed onto a smoother evolution spread out over a wider temperature range.[11,13,15] Over a temperature interval as large as HT-LT=40 K, however, one can consider that the greatest part of the valence change is completed.[11,13,15] Therefore, we adopt hereafter the approximation that [$Pr^{4+}$] goes from 0.13 (at LT) to 0 (at HT), allowing the analysis to be focused on the variation of [$Co^{3+}$ HS] across the transition.

*3.2 Jump in susceptibility around T\**

Let us first address the magnetism of the various cations at play in this system. The paramagnetic responses of $Pr^{3+}$ and $Pr^{4+}$ (isoelectronic to $Ce^{3+}$) depart only slightly from a standard Curie behavior, in such a way that their magnetism can be properly described by a temperature-dependent effective moment.[23-25] We use the values $\mu_{eff}(Pr^{3+})$ = 3.38 $\mu_B$ and $\mu_{eff}(Pr^{4+})$ = 2.48 $\mu_B$, derived from the local slopes around $T^*$ of the $1/\chi$ vs. $T$ curves given by Sekizawa *et al.*, that are modelizations of data measured in the series $PrSc_{1-x}Mg_xO_3$ (i.e., with Pr cations at the A sites of a perovskite structure).[26] For $Sm^{3+}$, the mixing effects between neighboring multiplets is more pronounced, leading to a substantial temperature-independent-paramagnetic (TIP) term, which can be dominant in the temperature range of present interest.[23-25,27] From a coupled investigation of $SmCoO_3$ and $GdCoO_3$, Ivanova *et al.* showed that the paramagnetism of $Sm^{3+}$ (on the A site of a perovskite structure), in the range 20-320 K, can be described by the combination of a small effective moment, $\mu_{eff}(Sm^{3+})$ = 0.47 $\mu_B$, and a large TIP term, $\chi_0(Sm^{3+})$ = 1.4 $10^{-3}$ emu/mol.[28] For $Co^{4+}$ IS and $Co^{3+}$ LS , we consider the spin-only values of the effective moments: $\mu_{eff}(Co^{4+}IS) = 3.87\ \mu_B$ and $\mu_{eff}(Co^{3+}LS) = 0\ \mu_B$. The case of $Co^{3+}$ HS is more complex, as illustrated by intense discussions about the transition at $T_1 \sim 100$ K in LaCoO$_3$.[16-20] The starting point of the controversy is that the $^5T_{2g}$ multiplet of the HS state is split by the spin-orbit coupling into a triplet, a quintet and a septet, with the triplet at the bottom.[16,20,29,30] There is currently a growing consensus to state that one must consider the excitation to this lowest triplet ($J'=1$) when addressing the population of the HS state around $T_1$ (the same should hold true around $T^*$ in our case). The effective Landé factor associated to this triplet ($g_{J'}$) was estimated from various techniques, leading to values in the range 3.2-3.5.[20,29,31,32] In the present study, we consider $\mu_{eff}(Co^{3+}HS) = 4.74\ \mu_B$, previously adopted by Knížek et al. for the description of LaCoO$_3$ around $T_1$, and which corresponds to the intermediate value $g_{J'} = 3.35$ ($g_{J'}\sqrt{J'(J'+1)} = 4.74$).[33]

To evaluate the paramagnetic susceptibility of Pr$_{0.49}$Sm$_{0.21}$Ca$_{0.3}$CoO$_3$ at each side of the transition, one must include a Curie-Weiss temperature to account for the magnetic interactions. The $1/\chi$ vs. $T$ plot displayed in Fig. 2 exhibits a clear Curie-Weiss behavior above $T^*$, from which is deduced $\theta_{CW} \sim -55$ K (over the range 110-150 K). This value can safely be adopted for HT=110 K, and we consider, in first approximation, that it also remains valid down to LT. The global susceptibility consists of the TIP term coming from Sm$^{3+}$, combined with the Curie-Weiss contributions from all the magnetic species, i.e.

$$\chi(T) = [Sm^{3+}] \times \chi_0(Sm^{3+}) + N_a\mu_{eff}^2 / [3k_B(T - \Theta_{CW})] \quad , \qquad (2)$$

where $N_a$ is the Avogadro number, $k_B$ the Boltzmann constant, and with

$$\mu_{eff}^2 = [Pr^{3+}] \times \mu_{eff}^2(Pr^{3+}) + [Pr^{4+}] \times \mu_{eff}^2(Pr^{4+}) + [Sm^{3+}] \times \mu_{eff}^2(Sm^{3+}) +$$
$$[Co^{4+}IS] \times \mu_{eff}^2(Co^{4+}IS) + [Co^{3+}HS] \times \mu_{eff}^2(Co^{3+}HS).$$

Taking into account the chemical formulation and relationships 1(a-c), this total effective moment reads

$$\mu_{eff}^2 = 0.49\,\mu_{eff}^2(Pr^{3+}) + 0.21\,\mu_{eff}^2(Sm^{3+}) + 0.3\,\mu_{eff}^2(Co^{4+}IS)$$
$$+ [Pr^{4+}] \times [\mu_{eff}^2(Pr^{4+}) - \mu_{eff}^2(Pr^{3+}) - \mu_{eff}^2(Co^{4+}IS)] + [Co^{3+}HS] \times \mu_{eff}^2(Co^{3+}HS).$$

Considering $[Pr^{4+}]_{LT} = 0.13$ and $[Pr^{4+}]_{HT} = 0$, the only parameter left to account for the experimental values of $\chi(LT)$ and $\chi(HT)$ is $[Co^{3+}\ HS]$. The values $z_1 = [Co^{3+}HS]_{LT}$ and $z_2 = [Co^{3+}HS]_{HT}$ allowing Eq. (2) to fit to the data are found to be $z_1 = 0.025$ and $z_2 = 0.395$, which correspond to the calculated $\chi$ values shown by diamonds in Fig. 2. Then, approximating the temperature dependence of $[Pr^{4+}]$ and $[Co^{3+}\ HS]$ between LT and HT by a phenomenological (rounded) step-function,[34] one can reasonably account for the variation of $\chi$ across the transition, as shown by the solid line in Fig. 2.

Let us now discuss the uncertainty related to the $\theta_{CW}$ value. First, we assumed the same $\theta_{CW}$ at LT and HT, while it might be expected to be lower (in absolute value) below $T^*$ when the $Co^{3+}$ are mainly LS (i.e., nonmagnetic). However, it turns out that $\theta_{CW}$ cannot be changed so much to keep Eq. (2) able to account for the $\chi(LT)$ value. Even if one considers $z_1 = 0$, $\theta_{CW}(LT)$ derived from Eq. (2) is found to be -47 K, i.e., not much smaller (in absolute value) than the -55 K used in the above analysis. It remains, however, that this uncertainty in $\theta_{CW}(LT)$ has an impact on $z_1$, whose value should thus be considered to be within the range 0-0.025. In other respect, the experimental uncertainty in $\theta_{CW}(HT) \approx -55$ K derived from $1/\chi$ vs. $T(>T^*)$ is estimated to be ± 5 K; one found that using -50 K or -60 K in Eq. (2) leads to values of $z_2$ equal to 0.37 or 0.42, respectively. Accordingly, one must consider that $z_2 = 0.395$ ± 0.025.

*3.3 Jump in entropy around T\**

The shift existing between the linear extrapolations of the $C(T)$ curve at each side of the transition in Fig. (1) is most likely ascribable to a variation in the lattice contribution. Indeed, a positive step in $C_{lattice}$ when crossing $T^*$ upon warming is expected owing to the expansion of the unit-cell volume ($V_u$) that is systematically present at the VSST.[1,2,6,9,11] The size of this change in $V_u$ was found to be ~ +1.5% in all these studies, either in $Pr_{0.5}Ca_{0.5}CoO_3$ or in mixed rare-earth compositions $(Pr_{1-y}Ln_y)_{1-x}Ca_xCoO_3$. Tsubouchi et al. (in $Pr_{0.5}Ca_{0.5}CoO_3$)[1] and Hejtmánek et al. [in $(Pr_{1-y}Y_y)_{0.7}Ca_{0.3}CoO_3$ ($y = 0.075, 0.15$)][8] previously pointed out a positive shift in the profile of the $C(T)$ curve across the transition.[1,8] In the present case, one can observe that the size of the shift in $C(T)$, estimated to be ~ 5 J K$^{-1}$mol$^{-1}$ from Fig. (1), is in line with these previous studies.

To limit as far as possible the impact of such a variation of $C_{lattice}$ on our estimate of the entropy change between LT(=70 K) and HT(=110 K), we use a background consisting of a straight line connecting the points $C(70K)$ and $C(110 K)$ (bold line in Fig. 3). The resulting excess heat capacity defined as $C_{ex} = C - C_{back}$ is shown in a $C_{ex}/T$ vs. $T$ plot in the inset of Fig. 3. Figure 4 displays the corresponding excess entropy calculated by $S_{ex}(T) = \int_{LT}^{HT} [C_{ex}(T)/T]\, dT$. One observes that the variation in entropy between LT and HT reaches a value $\Delta S_{ex}$ = 5.4 J K$^{-1}$ mol$^{-1}$.

Let us now try to derive a general expression for the entropy change at the VSST. The entropy terms to be taken into account are of different nature : magnetic ($S_{mag}$), mixing ($S_{mix}$), and electronic ($S_{elec}$). Each of these contributions is specified below:

- $S_{mag}$ is equal to $R\ln\nu$, where $R$ is the gas constant and $\nu$ is the total degeneracy. This degeneracy that can be given either by $\nu=(2J+1)$ (in case of a $J$-multiplet obtained after introduction of spin-orbit coupling), or by $\nu = \nu_{spin} \times \nu_{orb} = (2S+1) \times \nu_{orb}$ (in a scheme where the spin and orbital degrees of freedom are handled separately, as most often done for 3$d$ cations).

- $S_{mix}$ is a configurational entropy. It originates from the fact that there are various spatial distributions in which can be arranged the coexistence of different valence or spin states.

- $S_{elec}$ is an entropy associated to the presence of bandlike charge carriers. It must be introduced owing to the metallic-like conduction observed above $T^*$ [see Fig. 1(c)].

As a general rule, estimating the entropy by the sum of different terms requires that they originate from independent degrees of freedom, which can be a tricky issue in strongly correlated systems. In the present case, it turns out that the three terms considered above are rather well separated: indeed, $S_{mag}$ refers to the intra-cationic degrees of freedom, while $S_{mix}$ refers to the degrees of freedom associated to the global spatial distribution of these cations, and finally $S_{elec}$ refers to the degrees of freedom of the charge carriers moving between these cations. To ensure the reliability of this approach, note that the cationic interchange associated to the motion of these charge carriers must be taken into account in the evaluation of $S_{mix}$ (see below). In these conditions, and even though this remains an approximation, it appears legitimate to consider hereafter that $\Delta S_{ex} = \Delta S_{mag} + \Delta S_{mix} + \Delta S_{elec}$.

The various entropy terms are calculated hereafter in J K$^{-1}$ mol(f.u.)$^{-1}$, leading to use for each cation their content per f.u. (noted in square brackets as in the previous sections). As done for $\chi(T)$, one assumes [Pr$^{4+}$]$_{LT}$=0.13 and [Pr$^{4+}$]$_{HT}$=0, which restricts the free parameters to $z_1$ = [Co$^{3+}$HS]$_{LT}$ and $z_2$ = [Co$^{3+}$HS]$_{HT}$. Since only Pr and Co are involved in the entropy changes (the absence of valence transition affecting Sm has been verified by XAS),[15] the global entropy change is the sum of five terms:

$\Delta S_{ex} = \Delta S_{mag}(\text{Pr}) + \Delta S_{mag}(\text{Co}) + \Delta S_{mix}(\text{Pr}) + \Delta S_{mix}(\text{Co}) + \Delta S_{ele}$. (3)

$\Delta S_{mag}$(Pr): For $Pr^{3+}$ and $Pr^{4+}$ around $T^*$, one can consider the full degeneracy of the ground-state multiplets, i.e. $\nu = 2J+1$ with $J(Pr^{3+}) = 4$ and $J(Pr^{4+}) = 5/2$, leading to $\Delta S_{mag}$(Pr)=R{(0.49 ln9)-(0.36 ln9+0.13 ln6)}, which simplifies to

$$\Delta S_{mag}(\text{Pr}) = R\{0.13(\ln 9 - \ln 6)\}. \tag{4}$$

$\Delta S_{mag}$(Co): $Co^{4+}$IS has the electronic configuration $t_{2g}^4 e_g^1$, yielding $S=3/2$ and thus $\nu_{spin} = 4$. As for the orbital degeneracy, the Tanabe-Sugano diagram of $3d^5$ shows that the lowest level associated to the IS state is a triplet ($^4T_{1g}$ in octahedral environment) separated from the upper level by about 1eV.[35] A similar situation holds for $3d^6$, which was previously used in the literature to associate a triple orbital degeneracy to $Co^{3+}$ IS.[16,17,30] In other respect, one can alternatively consider that the likely presence of a Jahn-Teller effect affecting the $e_g$ orbitals in the configuration $t_{2g}^4 e_g^1$ requires to take into account only the degeneracy within the $t_{2g}$ triplet, leading once again to $\nu_{orb}(Co^{4+}$ IS) = 3. This latter argument was also used in the literature on $Co^{3+}$ IS –which shows the same $e_g$ filling ($t_{2g}^5 e_g^1$)– to justify a triple orbital degeneracy.[18,36] In the end, one thus obtains that $\nu(Co^{4+}$ IS) = $\nu_{spin} \times \nu_{orb}$ =12. For $Co^{3+}$HS, the configuration $t_{2g}^4 e_g^2$ should, in principle, yields $\nu_{spin} = 5$ and $\nu_{orb} = 3$, leading to $\nu = 15$. Nevertheless, as previously discussed, there is a growing agreement to state that the spin-orbit induced splitting of the $Co^{3+}$ HS state must be taken into account at intermediate temperatures. Consistently with Section 3.2, we thus consider that the excitation of $Co^{3+}$ from LS to HS deals with the lowest triplet ($J'=1$) of the latter state, leading to $\nu(Co^{3+}HS) = 2J'+1 = 3$. Accordingly,

$$\Delta S_{mag}(\text{Co}) = R\{(z_2 - z_1)\ln 3 + 0.13 \ln 12\}. \tag{5}$$

$\Delta S_{mix}$(Pr): The Pr sites can host either $Pr^{3+}$ or $Pr^{4+}$ states. Keeping in mind that there are only 0.49 Pr per f.u., the corresponding mixing entropy per f.u. is

$$S_{mix}(\text{Pr}) = -R\{[Pr^{3+}]\ln[Pr^{3+}] + [Pr^{4+}]\ln[Pr^{4+}] - 0.49\ln 0.49\}.$$

Since there is no mixing effect at HT ($[Pr^{4+}]_{HT}=0$), $\Delta S_{mix}$ is just the opposite of the above expression evaluated at LT:

$$\Delta S_{mix}(\text{Pr}) = +R\{0.36\ln 0.36 + 0.13\ln 0.13 - 0.49\ln 0.49\}. \tag{6}$$

Δ$S_{mix}$(Co): There are three cobalt species ($Co^{4+}$IS, $Co^{3+}$LS, $Co^{3+}$HS), whose sum of contents is equal to 1 per f.u.. As long as these cations are well distinguishable, the mixing entropy reads as

$$S_{mix}(Co) = -R\left\{[Co^{4+}IS]\ln[Co^{4+}IS] + [Co^{3+}HS]\ln[Co^{3+}HS] + [Co^{3+}LS]\ln[Co^{3+}LS]\right\}.$$

Above $T^*$, however, a metallic-like behavior sets in, as shown in Fig. 1(c). This corresponds to an easy mobility of charge carriers which can be ascribed to the interchange of one $e_g$ electron between $Co^{4+}$IS ($t_{2g}^4 e_g^1$) and $Co^{3+}$HS ($t_{2g}^4 e_g^2$).[37] As a consequence, one should just consider in this case the mixing between $Co^{3+}$LS and the two other species taken together:

$$S'_{mix}(Co) = -R\left\{([Co^{4+}IS]+[Co^{3+}HS])\ln([Co^{4+}IS]+[Co^{3+}HS]) + [Co^{3+}LS]\ln[Co^{3+}LS]\right\}.$$

Therefore, to evaluate the mixing entropy change around $T^*$, we consider $S_{mix}$ below $T^*$ and $S'_{mix}$ above $T^*$. The resulting entropy change is

$$\Delta S_{mix}(Co) = +R\left\{\begin{array}{l}0.17\ln 0.17 + z_1\ln z_1 + (0.83-z_1)\ln(0.83-z_1) \\ -(0.3+z_2)\ln(0.3+z_2) - (0.7-z_2)\ln(0.7-z_2)\end{array}\right\} \quad (7)$$

Δ$S_{elec}$: The estimate of such an electronic term is always delicate. For an insulating-to-metallic (IM) transition (as occurring at $T^*$), it is customary to approximate it by

$$\Delta S_{elec}(T^*) = \gamma_{metal} T^*, \quad (8)$$

where $\gamma_{metal}$ is linked to the density of states at the Fermi surface. It must be emphasized that this $\gamma_{metal}$ corresponds to $T > T^*$, so it cannot be simply identified with $\gamma$ (the linear coefficient of electronic specific heat) commonly derived from the extrapolation of $C/T$ vs. $T^2$ to $T\to 0$. Actually, as pointed out by Knížek et al.,[38] the $\gamma_{metal}$ relevant to Eq. (8) is expected to be significantly smaller than $\gamma$ derived from heat capacity at low-$T$. The resistivity curve $\rho(T)$ in Fig. 1(c) points to a striking similarity between the behavior observed in $(Pr_{0.7}Sm_{0.3})_{0.7}Ca_{0.3}CoO_3$ above $T^*$ and that found in $LaCoO_3$ above the IM transition (around $T_2 \sim 450$ K), i.e., a resistivity $\rho \sim 2$ mΩ.cm showing a slight decrease as $T$ is increased.[36,39,40] The same features are actually observed in all the $(Pr_{1-y}Ln_y)_{1-x}Ca_xCoO_3$ compounds showing the VSST.[1,2,4-6,8] Addressing the issue of $\gamma_{metal}$ in $LaCoO_3$ above the IM transition, Stølen et al. suggested a value 5 mJ $K^{-2}$ $mol^{-1}$,[16] while Tachiban et al. estimated that it should lie in the range 10-20 mJ $K^{-2}$ $mol^{-1}$.[41] On their side, Knížek et al. experimentally derived a value 5-6 mJ $K^{-2}$ $mol^{-1}$ in the metallic-like region of the closely related $PrCoO_3$ and $NdCoO_3$ compounds.[38] On the basis of these results, we will consider that the variation in electronic entropy between LT=70 K (insulating-like regime) and HT=110 K (metallic-like regime) in $(Pr_{0.7}Sm_{0.3})_{0.7}Ca_{0.3}CoO_3$ can be approximated by $\Delta S_{elec} = \gamma_{metal} \times HT$, with $\gamma_{metal}$ being in the range 5 – 10 mJ K$^{-2}$ mol$^{-1}$.

One can now evaluate each term of Eq. (3). The two contributions from Pr are $\Delta S_{mag}$(Pr) = +0.438 J K$^{-1}$ mol$^{-1}$ and $\Delta S_{mix}$(Pr) = -2.36 J K$^{-1}$ mol$^{-1}$. The latter value is negative since there is an increased configurational entropy below $T^*$ resulting from the appearance of a mixed valency. For the Co contributions, using the values $z_1$=0.025 and $z_2$=0.395 derived from the $\chi(T)$ analysis, one obtains $\Delta S_{mag}$(Co) = +6.06 J K$^{-1}$ mol$^{-1}$ and $\Delta S_{mix}$(Co) = +0.39 J K$^{-1}$ mol$^{-1}$. Finally, depending on the adopted $\gamma_{metal}$ value, $\Delta S_{elec}$ is comprised between +0.55 et +1.1 J K$^{-1}$ mol$^{-1}$. Combining all these values, $\Delta S_{ex}$ is expected to be between 5.08 and 5.63 J K$^{-1}$ mol$^{-1}$.[42] It turns out that the experimental value, 5.4 J K$^{-1}$ mol$^{-1}$, is well within this range. Therefore, one can conclude that the variations in susceptibility and entropy around $T^*$ can be both accounted for by the same values of Co$^{3+}$ HS content, a result which supports the relevancy of the LS/HS scenario to the VSST.

*3.4 Comparison to the literature*

About the jump in $\chi(T)$ at a VSST, the only quantitative analysis similar to ours was performed by Fujita *et al.* on $(Pr_{1-y}Sm_y)_{1-x}Ca_xCoO_3$ compounds.[4] This study was conducted within the framework of a LS-to-IS transition for Co$^{3+}$, and concluded to the occurrence of a partial transformation. This result, however, did not take into account the presence of a valence transition since it was not yet revealed in 2005. There are more numerous studies in the literature about the entropy change at the transition, determined from $C(T)$ data.[1,7,8] The procedure followed for the derivation of a $\Delta S_{ex}$ value was in all cases similar to ours. In $Pr_{0.5}Ca_{0.5}CoO_3$, Tsubouchi *et al.* found $\Delta S_{ex} \approx 4.7$ J K$^{-1}$ mol$^{-1}$,[1] while Kalinov *et al.* reported $\Delta S_{ex}$ values in the range 2.5-2.6 J K$^{-1}$ mol$^{-1}$ for $(Pr_{1-y}Eu_y)_{0.7}Ca_{0.3}CoO_3$ with $y$ = 0.22-0.26.[7] In both studies, the $\Delta S_{ex}$ values were claimed to be consistent with the $\Delta S_{mag}$ of Co$^{3+}$ expected for an LS-to-IS transition, when considering only the spin degeneracy. It works quite well indeed for $Pr_{0.5}Ca_{0.5}CoO_3$, since $0.5R\ln3 \approx 4.6$ J K$^{-1}$ mol$^{-1}$, but Kalinov *et al.* had to involve the additional assumption that the amount of Co$^{3+}$ undergoing the transition is given by the Co$^{4+}$ content (i.e., $0.3R\ln3 \approx 2.7$ J K$^{-1}$ mol$^{-1}$). Beyond the nature of the Co$^{3+}$ spin-state, these analysis are very different from ours in that they neglected the valence change, the mixing effects, and the presence of an electronic term.

In (Pr$_{1-y}$Y$_y$)$_{0.7}$Ca$_{0.3}$CoO$_3$ compounds, Hejtmánek *et al.* reported $\Delta S_{ex}$ equal to 2.17 and 4.78 J K$^{-1}$ mol$^{-1}$, for $y$ = 0.075 and 0.15, respectively.[8] These values were analyzed in a way similar to the present study, i.e., combining magnetic, mixing and electronic contributions.[43] Maryško *et al.* assumed an LS-to-IS transition for Co$^{3+}$, as in the other studies of the literature, but reported that the entropy change at $T^*$ is compatible with a total transformation from LS to IS if one takes into account the presence of mixing and electronic entropies.[43] It can be noted that this analysis still differs from ours on various issues, since, for instance, mixing was considered only for the Co cations and the Pr/Co valence changes were not taken into account.

## 4. Conclusion

In (Pr$_{0.7}$Sm$_{0.3}$)$_{0.7}$Ca$_{0.3}$CoO$_3$, which exhibits a VSST centered at $T^* \sim 90$ K, we determined the variation in susceptibility and in entropy between two references temperatures flanking the transition: 70 K (referred to as LT) and 110 K (referred to as HT). Using previous XAS results, we calculated expressions of $\chi(T)$ around $T^*$ and of $\Delta S_{ex}$ across $T^*$, in which the only unknown parameter was the content of Co$^{3+}$ HS. First, the comparison to the $\chi(T)$ data led to a change in the amount of Co$^{3+}$ HS that goes from ~0.025 to ~0.395 per f.u. between the two reference temperatures. Second, the inclusion of these values in the calculated $\Delta S_{ex}$ was found to yield a result consistent with the experimental one (~ 5.4 J K$^{-1}$ mol$^{-1}$). Our theoretical evaluation of $\Delta S_{ex}$ combined various contributions: magnetic terms accounting for the spin and orbital degrees of freedom; mixing terms related to the coexistence of various valence or spin states; and an electronic term associated to the metallic-like behavior observed above the transition. We considered –both in the magnetic and calorimetric analysis– that the excitation to the HS state of Co$^{3+}$ deals with the lowest triplet resulting from the spin-orbit splitting, leading to $\mu_{eff}$(Co$^{3+}$ HS) = 4.74 $\mu_B$ (effective g-factor ~ 3.35 ) and $S_{mag}$(Co$^{3+}$ HS) = $R$ ln3.

To sum up, we observed that the jump in $\chi(T)$ and the peak in $C(T)$ around $T^*$ are both compatible with a scenario where the VSST corresponds to a partial Co$^{3+}$LS to Co$^{3+}$HS transition, coexisting with Co$^{4+}$ in the IS state, and coupled with Pr$^{4+}$/Pr$^{3+}$ and Co$^{3+}$/Co$^{4+}$ valence changes.

**Figures captions**

Fig. 1: Temperature dependence of three physical properties around $T^*$ in $(Pr_{0.7}Sm_{0.3})_{0.7}Ca_{0.3}CoO_3$: (a) heat capacity; (b) magnetic susceptibility; (c) electrical resistivity. The solid lines in (a) correspond to phenomenological linear regimes at each side of the transition. The solid line in (b) displays the Curie-Weiss behaviour present at $T > T^*$. In each panel, the vertical dashed lines correspond to the reference temperatures LT (=70 K) and HT (=110 K).

Fig. 2: Reciprocal susceptibility at the transition: the circles are the experimental data; the solid diamonds correspond to the $[Co^{3+}$ HS] contents derived at LT and HT (see text); the bold solid line is calculated by approximating the temperature dependence of $[Co^{3+}$ HS] and $[Pr^{4+}]$ between LT and HT by a rounded step-function.[34] The thin solid line above $T^*$ is a Curie-Weiss law leading to $\theta_{CW}$ = -55 K.

Fig. 3: Heat capacity around the transition. The bold solid line between LT and HT is the background used for the derivation of the excess specific heat, $C_{ex}$. The inset shows $C_{ex}/T$ vs. $T$.

Fig. 4: Temperature variation of the excess entropy $S_{ex}$ (derived from $C_{ex}$) over the range LT – HT. It saturates to $\Delta S_{ex} \approx 5.4$ J K$^{-1}$ mol$^{-1}$.

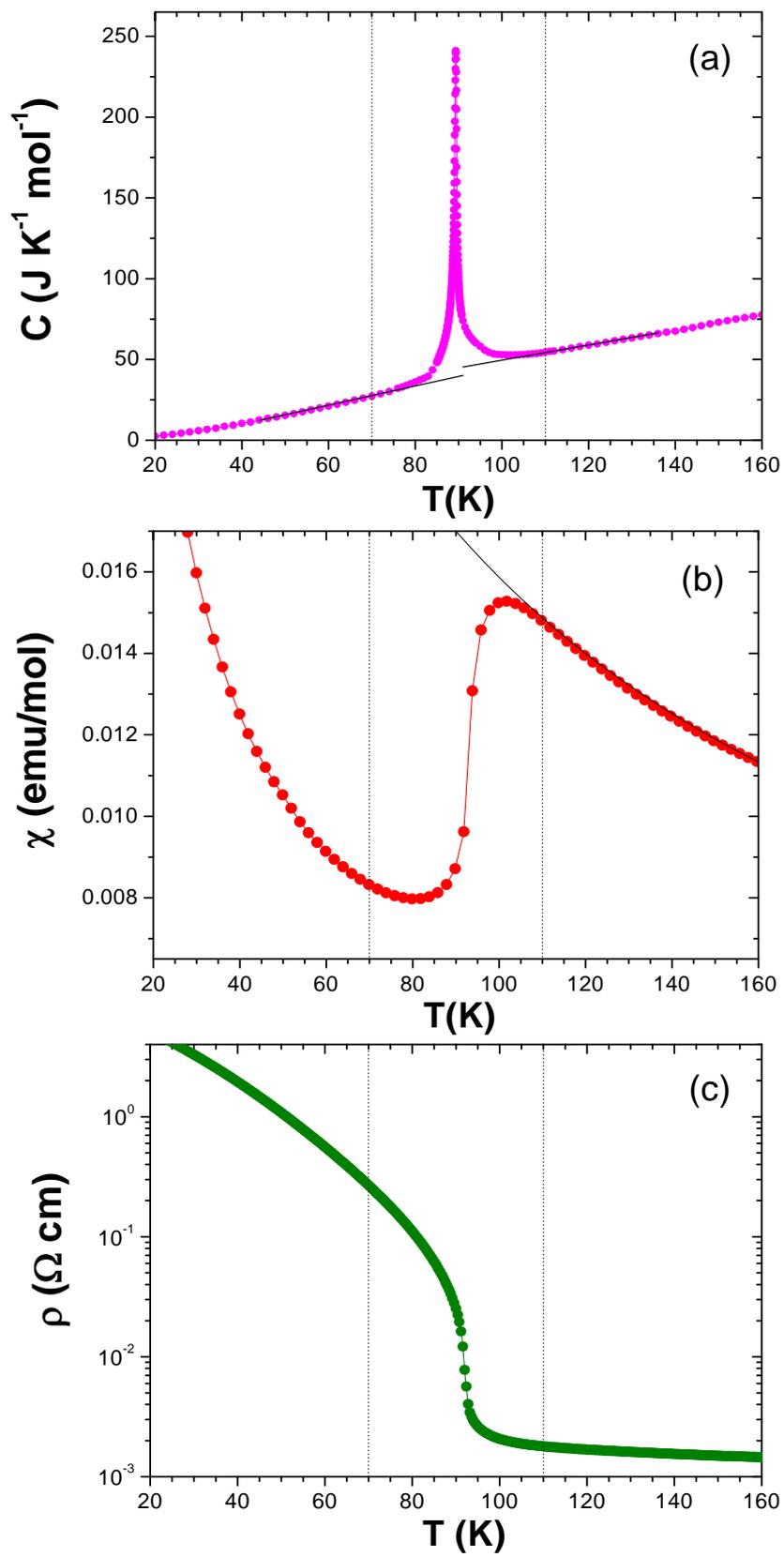

Figure 1
Hardy *et al.*

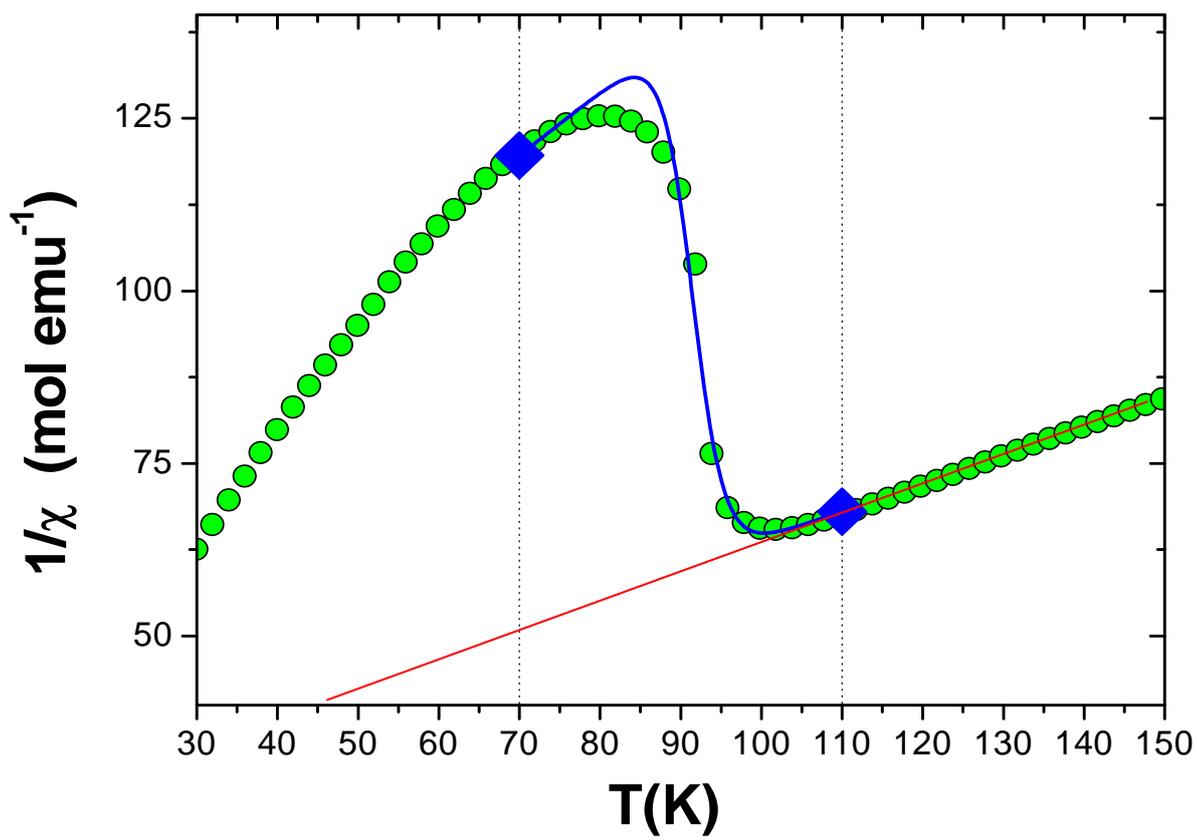

Figure 2
Hardy *et al.*

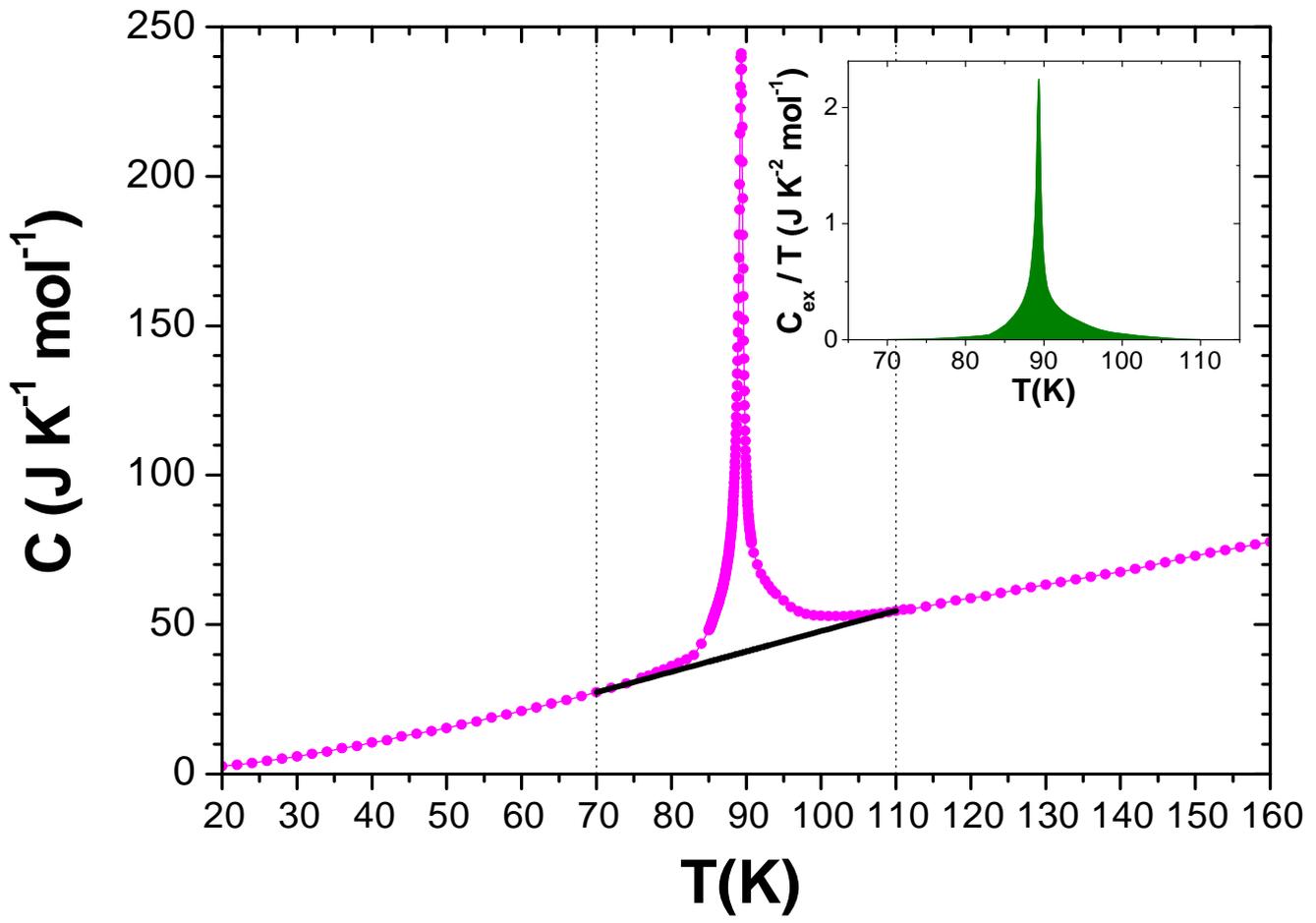

Figure 3
Hardy *et al.*

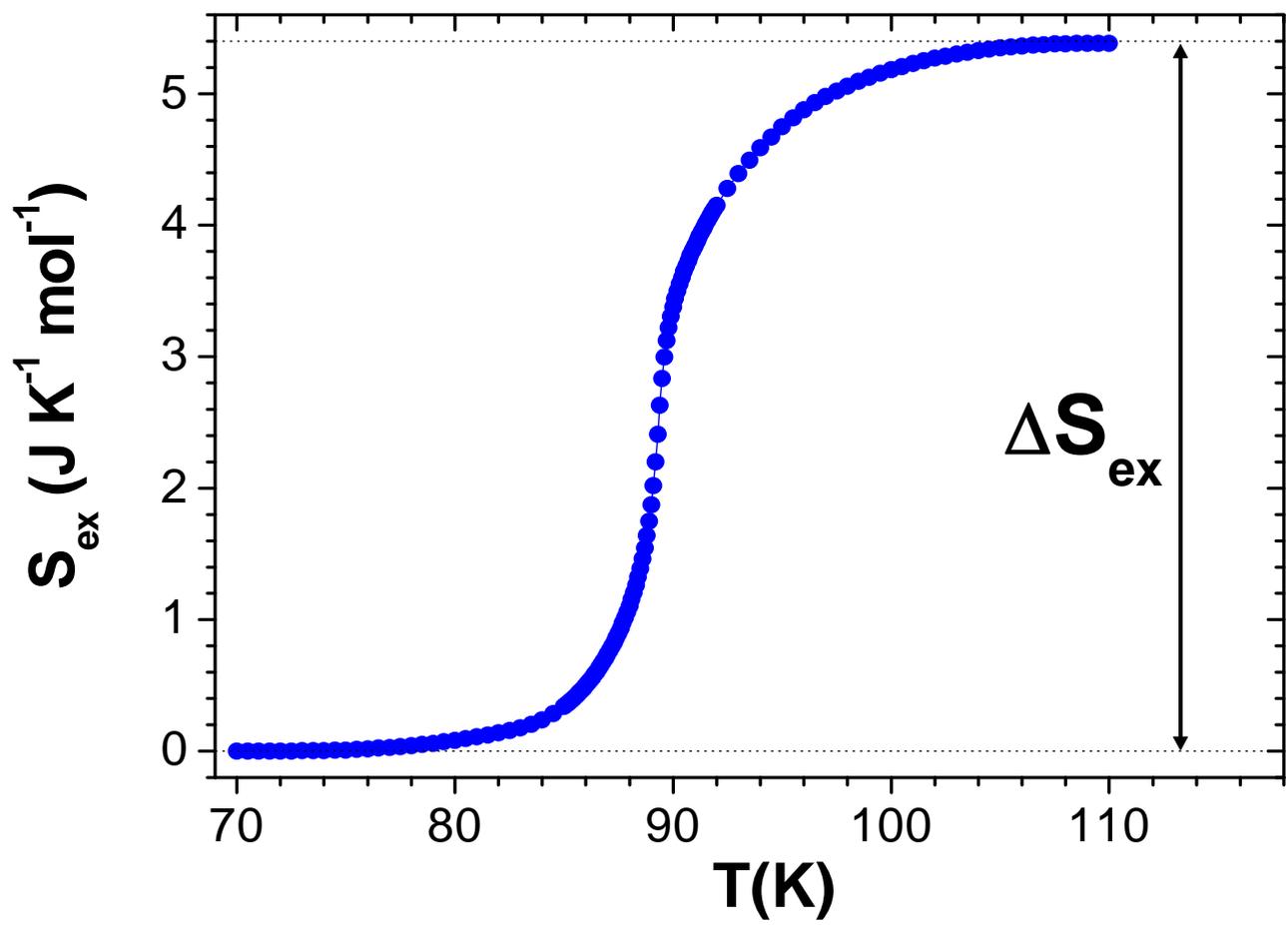

Figure 4
Hardy *et al.*